\documentclass[a4paper,reqno,11pt,oneside]{amsart}
\usepackage{amsfonts,amssymb,latexsym,xspace,epsfig,color}
\usepackage{amsmath,stmaryrd,xy}
\usepackage{a4wide}
\usepackage[small,bf]{caption}
\usepackage{enumitem}
\usepackage{float}
\usepackage{graphicx}
\usepackage{mathtools, nccmath}
\usepackage{cite}
\usepackage[utf8]{inputenc}
\usepackage[nodisplayskipstretch]{setspace}
\usepackage{mathtools}
\usepackage{comment}

\DeclarePairedDelimiter\floor{\lfloor}{\rfloor}

\bibdata{prob}
\bibstyle{alpha} 

\theoremstyle{plain}

\theoremstyle{definition}

\newcommand{\norm}[1]{\left\lVert#1\right\rVert}
\numberwithin{equation}{section}
\allowdisplaybreaks[1]

\newlist{lteo}{enumerate}{1}
\setlist[lteo,1]{wide,itemsep={6pt plus 4pt},font=\bfseries,label={(\arabic*)}}

\newlist{iteo}{enumerate}{1}
\setlist[iteo,1]{wide,itemsep=2ex,label={\upshape(\roman*)}}

\makeatletter
\@namedef{subjclassname@2020}{%
  \textup{2020} Mathematics Subject Classification}
\makeatother

\title[Application of Information Theory in  Rumor Spreading  Modeling ...]{Application of Information Theory in  Rumor Spreading  Modeling Considering Polarization in Complex Networks}

\author{Patrick Oliveira de Paula, Alejandra Rada and Catalina R\' ua}

\date{}

\address{
\newline
 Alejandra Rada and Patrick Oliveira de Paula 
\newline 
UFABC - Centro de Matem\'atica, Computa\c{c}\~ao e Cogni\c{c}\~ao
\newline
Avenida dos Estados, 5001- Bangu - Santo Andr\'e - S\~ao Paulo, Brasil.
\newline
e-mails:   alejandra.rada@ufabc.edu.br, patrick.oliveira@aluno.ufabc.edu.br
\newline
\vspace{0.2cm}
\newline
Catalina R\'ua Alvarez
\newline
Universidad de Nari\~no
\newline
Cr. 33 No. 5 - 121 - Pasto, Colombia
\newline
e-mail: catalina.rua@udenar.edu.co
}

\thanks{This work has been partially supported by FAPESP Grant 2018/22972-8}

\begin{document}

\begin{abstract}
  It is proposed a model that makes use of the Entropy concept and the Jensen-Shannon information divergence function to simulate computationally the dissemination of opinions on a Barab\'asi-Albert (BA) scale-free network. The simulation considers individual memory, ideological proximity between individuals, and distortion influenced by  polarization. 
\end{abstract}

  \maketitle

\section{Introduction}
Information exchange and opinion sharing are essential aspects of human social interaction. The dissemination of information and rumors can have significant effects on society and its political and economic structures. The information can influence public opinion about political conjunctures \cite{galam2003modelling}, impact financial markets \cite{kosfeld2005rumours, kimmel2004rumors} and create social tensions and unnecessary generalized anxiety \cite{grein2000rumors, thomas2007lies} about possibly false public issues. Recently, the spread of false rumors has hampered the challenge of solving the Covid-19 pandemic, increasing public hesitation in the face of new vaccines  \cite{carrieri2019vaccine} and favoring the adoption of interventions without scientific support in terms of their effectiveness for the treatment of Covid-19. Models that study and help to have some control over the dissemination of rumors could be crucial to combat the damage caused by misinformation and, above all, fake news.

The standard model for rumors was introduced by Daley and Kendall in the mid-1960's. In a communication published by Nature, \cite{DKnature}, they proposed a simple mathematical model for rumor transmission which appeared as an alternative to the well-known susceptible-infected-removed epidemic model (SIR). This model was modified  by  Maki and Thompson (MT) in 1973, \cite{maki1973mathematical}, changing the rules of the rumor transmission. After this, there were generalizations of those models considering the dynamic of the rumor dissemination process and the population structure. Models obtained by altering the process dynamic were worked on by Sudbury \cite{sudbury1985proportion}, who was interested in the proportion of the population that never heard the rumor;  Lebensztayn \textit{et al}. \cite{LEBENSZTAYN2011517} assumed that each informant ceases to propagate the rumor after contacting either another informant or someone who actually knew the rumor; Belen \textit{et al}. \cite{Belen2011} studied the continuous-time MT model; Arruda \textit{et al}. \cite{Arruda2015} proposed a competition model in which informants try to transmit information, while those that knew the rumor try to hold on to it and finally, Rada \textit{et al}. \cite{Rada2021} formulated a generalization for the MT model, by assuming that each ignorant becomes a spreader only after hearing the rumor a certain number of times. Models obtained by altering the structure of the population were studied by Coletti \textit{et al}. \cite{Coletti2012} in 2012, where the rumor process was analyzed when the population is represented by the $d$-dimensional hypercubic lattice; Comets \textit{et al}. \cite{Comets2013} modeled the transmission of information in the Erd\"{o}s-R\'enyi  random graph and Gallo \textit{et al}. \cite{Gallo2014} studied rumors in $\mathbb{N}$ using renewal processes.

With the development of complex networks, the dynamic of spreading rumors gained the topology of social networks. Moreno \textit{et al}. \cite{moreno2004efficiency} investigated the stochastic version of the MT model in scale-free networks using Monte Carlo simulations; Nekovee \textit{et al}.  \cite{Nekovee2008} in 2008, improved the MT model to describe the dynamics in complex social networks. In 2013, Zhao \textit{et al}. \cite{Zhao2013} studied a rumor dissemination model considering a forgetting rate in a small-world network. In 2017, Agliari \textit{et al}. \cite{Agliari2017} studied the phase transition of the MT model also for a small-world network. Computational and simulation techniques were applied in these models.

Until now, a limited number of articles have explored on the dissemination of rumors using information theory. In 2017, Wang \textit{et al}. \cite{wang2017rumor} proposed a model for the dissemination of a rumor where the informant can distort the message with a probability that depends on empirical entropy and the individual who receives the message has the possibility to accept it or not depending on the number of connections of the informant. The results were obtained through simulations in a Barab\'asi-Albert scale-free network (BA scale-free network). In 2019, also Wang \textit{et al}. \cite{wang2019progressive} modified the last work,  adding  polarization's effects. The model considers, besides the possibilities of distortion and acceptance, incorporates the possibility of information being distorted either positively or negatively. In 2022, Jain \cite{Jain2022} proposed a model which controlling COVID-19 rumors through the power of opinion leaders based on entropy. Also in 2022, Lei and Cheong \cite{LeiCheong} using a Local Structure Entropy (LSE) approach for proposing a method, based on the Taslli entropy,  to explore the impact of removing nodes on the network. Recently in 2024,  Wang \textit{et al}. \cite{wangG2024}  proposed a method from the perspective of information dissemination for finding influential nodes based on Kullback–Leibler divergence model within the neighborhood.

In this paper is proposed a rumor's dissemination model that considers the ideological proximity and polarization distortion tendency between individuals. The model for propagating opinions considers the memory capacity of individuals who are going to store, as information, sequences of symbols of fixed length.  The information in the memory of each individual allows the application of empirical entropy as a measure of heterogeneity of opinions. 
Furthermore, a measure of information divergence is used, in particular the Jensen-Shannon measure, to quantify the ideological proximity between individuals, identified as the weight of network connections. This proximity starts to be considered in the calculation of the probability of acceptance of information. To integrate potential polarization effects in the propagation of information, it is used a function to quantify the polarity of information and a probability of biased distortion, depending on the individual's polarization tendency (positive or negative) of the propagated information.

The paper is organized as follows. In Section \ref{S:model}, it is defined the model and the algorithm used in the simulations. In the section 3 it is found the results and in the final section it is made a discussion. 

\section{Model}\label{S:model}

In this section, it is present the rumor propagation model on a non-oriented, weighted graph of Barab\'asi-Albert of size $N$, denoted by $G$. In it,  the individuals are the nodes and their social connections are the edges. Analogously to \cite{wang2017rumor,wang2019progressive}, the spread of information occurs in three consecutive phases: \textit{information spreading, information acceptance, information update}.

\subsection{Information spreading}
In this phase, the information piece will be a binary string of length $s$. For instance, if $s=5$, the information piece might be ``10001". The total of all different information pieces is $2^s$ and they compose the alphabet $\mathcal{A}$, the set of all possible information pieces.

Each individual \(u \in v(G) \), where $v(G)$ is the set of nodes, will be assigned an ensemble \((L_u, P_u, X_u)\). $L_u$ is a list of $\mu$ information pieces representing the memory capacity of $u$. Counting the empirical frequencies of binary string inside $L_u$ makes it possible to define a probability distribution $P_u$, which describes the outcomes of a random variable $X_u$ with sample space given by $\mathcal{A}$. Hence, for an individual $u$ and a given $x \in \mathcal{A}$,
\begin{equation*}
    P_u (x) = P_u \{X_u = x\} = f_{x},
\end{equation*}
where $f_{x}$ is the frequency of the information piece $x$ in the list $L_u$. 
For each individual $u$ it is defined the empirically Shannon entropy,
\begin{equation*}
   H_u = H(P_u) = - \sum_{x \in L_u} f_x \log_2 {f_x}.
\end{equation*}

The entropy increases as the information distribution of the individual's memory tends to a Uniform probability distribution (i.e., $f_x \to 1 / 2^s$), and decreases as single information predominates ($f_x \to 0$). The average entropy for the population will be
\begin{equation*}
    \overline{H} = \frac{1}{|v(G)|} \sum_{u \in v(G)} H_u.
\end{equation*}

Entropy provides a measure of the heterogeneity of information that each individual holds in their memory. Using this measure, the individual can distort the information to be transmitted with the following probability,
\begin{equation}\label{delta}
    \delta_u = \frac{1}{\exp{\left[\left(\frac{H_{\text{max}} - H_u}{H_{\text{max}}}\right) \cdot \kappa\right] + 1}},
\end{equation}
where $\kappa \geq 0$ is a \textit{conservation factor}, a control parameter of the general tendency for information distortions. In this model, it is understood that the information transmission between nodes defines a noisy communication channel, where each binary string is selected for transmission bit by bit. Therefore, $\delta_u$ defines a probability for bit-wise distortion.

The use of information stored in memory allows to consider the effects of polarization or bias on the  information to be transmitted. For any information piece $x = b_1 b_2 \ldots b_s$, where $b_i \in \{0, 1\}$ for each $i = 1, \ldots, s$, the polarity of $x$ is defined as the mean $\pi(x) = \frac{1}{s} \sum_{i = 1}^{s} b_i$, and it naturally follows a polarity measure for an individual $u$ by $\pi_u = \frac{1}{\mu} \sum_{x \in L_u} \pi(x)$. The individual tendency of polarization will play an important role as a distortion bias over the communication channel between $u$ and their neighbors, which becomes a binary asymmetric channel instead of the symmetric channel.  

It is defined the probability of distortion due to polarization tendency as
\begin{equation}\label{xi}
    \xi_u = \frac{1}{\exp{[\lambda]} + 1}.
\end{equation}
The intensity with which this polarization happens will be governed by the \textit{polarization resistance} $\lambda >0 $.

The individuals in the network will be randomly grouped according to their bias. Taking two parameters $\alpha, \omega \in [0, 1]$, such that $\alpha + \omega \leq 1$, it is possible to partition the set $v(G)$ in three groups, $V_{\downarrow}$, $V_{\uparrow}$ and $\overline{V}$, where 

\begin{eqnarray*}
|V_{\downarrow}| &=& \floor{\alpha \cdot |v(G)|},\\
|V_{\uparrow}| &=& \floor{\omega \cdot |v(G)|} \quad \quad \text{and} \\
|\overline{V}| &=& |v(G)|-(|V_{\downarrow}|+|V_{\uparrow}|),
\end{eqnarray*}
and each partition is filled randomly and uniformly. The set $V_{\downarrow}$ contains the group of individuals who lean towards downward polarization, that is, having the tendency to change 1 for 0. $V_{\uparrow}$ the individuals leaning towards upward polarization, in other words, they have the tendency to change 0 for 1 and $\overline{V}$ the neutral ones, that is, they change 1 for 0 and vice versa without preferences.

Let $E$ be the set of edges of the graph $G$. The algorithm  selects each edge $\langle u, v \rangle \in E$ exchanging information pieces between $u$ and $v$, that is, it takes the information $x$ and $y$ from $X_u$ and $X_v$,  according to the distributions $P_u$ and $P_v$ and then, they are exchanged between $u$ and $v$, respectively. 

\subsection{Information acceptance}
By attributing weights to the  edges of the graph, it is modeled the ideological proximity between individuals, serving as a base to quantify the propensity of information acceptance between individuals. With this purpose in mind, it is applied the Jensen-Shannon divergence measure $(JSD)$, or information radius, over the distributions $P_u$ and $P_v$ of two neighbors $u$ and $v$ to compare the information content of their memories:
\begin{equation*}
    \begin{split}
        JSD(P_u; P_v) &= \frac{1}{2} D(P_u; M) + \frac{1}{2} D(P_v; M), \\
        &= H(M) - \frac{H(P_u) + H(P_v)}{2},
    \end{split}
\end{equation*}
where $M = (P_u + P_v) / 2$ and $D(\cdot ; \cdot)$ is the KL divergence (see \cite{Lin,Endres}). The $JSD$ function has its image contained in the interval $[0, 1]$, being a measure of divergence and $JSD(P, Q) = 0$ if, and only if, $P = Q$. It is defined the proximity between the individuals $u$ and $v$ as the function $\mathcal{S}(P_u, P_v)$ given by 
\begin{equation*}
    \mathcal{S}(P_u; P_v) = 1 - JSD(P_u; P_v).
\end{equation*}

The strength of connections between an individual $u$ and its neighbors can be measured by the function $S(\cdot; \cdot)$. If $k_u$ is the degree of $u$ and $N(u)$ is the set of neighbors of $u$, it is hold that
\begin{equation*}
    \sigma(u) = \sum_{w \in N(u)} \mathcal{S}(P_u ; P_w) \leq k_u.
\end{equation*}

The value $\sigma(u)$ gives a measure of social connectivity for an individual $u$ since it considers the ideological similarity between $u$ and its neighbors. For a given $v \in N(u)$, the acceptance probability of information propagated by $u$ depend on the proximity between $u$ and $v$ and the relative popularity between $u$ and its neighbors. This relative popularity is represented by the term
\begin{equation*}
    \frac{\sigma(u)^{\gamma}} { \max_{w \in N(v) \cup \{v\}}{\sigma(w)^{\gamma}}}.
\end{equation*}
If $\gamma > 0$ and the relative popularity among neighbors is equal to $1$ meaning that $u$ is the most popular neighbor of $v$, and decays if any other neighbor is more influential than $u$. For $\gamma <0$ the reasoning is analogous. 

The probability of $v$ accepting information from $u$ it is denoted by $\eta_{u \to v}$ and it is defined as 
\begin{equation}
    \eta_{u \to v} = \frac{2}{\mathcal{S}(P_u; P_v)^{-1} + \left(\frac{\sigma(u)^{\gamma}}{\max_{w \in N(v) \cup \{v\}}{\sigma(w)^{\gamma}}}\right)^{-1}}.
\end{equation}

Since the harmonic mean is strongly influenced by minimum values, we decided to use it to penalize any asymmetry between proximity and  relative popularity among neighbors. The parameter  $\gamma$ is called the \textit{confidence factor} and represents the global preference towards more ($\gamma > 0$) or less ($\gamma < 0$) influential individuals.

\subsection{information update}
The network will be initialized with the memories of all individuals randomly filled. 
The updating of memory banks occurs synchronously across individuals: at every time-step $t$ of the process, all individuals try to disseminate to the neighbors the information whose probability is the highest, and then all individuals decide synchronously whether to accept or reject new information received from their neighbors, after which, the memory banks are updated and as a result, a new set of values will be available in memory at a time $t+1$. For example, if an individual $u$ has five neighbors,  $u$ will try to spread one piece of information to each of them, potentially distorted and polarized, and receive five pieces of information in return. Some of them will be rejected and others accepted.  Assuming that $u$ accepts only two pieces of information, those pieces will enter on $u$'s memory bank. If $u$ was already at maximum memory capacity, the two oldest pieces in the bank will be forgotten.

\section{Results}

The model is studied in a previously created BA graph, denoted by $G$. To create the graph, it was started $m_0 = 5$ connected nodes and new nodes were added, one at a time. Each new node was connected to  $m = 2$ existing nodes with a probability proportional to the degree that the existing nodes already have. The final number of nodes was defined as $N = 500$, and the memory size of each node was defined as $\mu = 256$. All results were obtained on the same generated
graph.

The memory of every individual works as a \textit{first-in, first-out} (FIFO) line, and it starts with two possible initial distributions: the Binomial distribution or the Uniform distribution. Let $L_u$ be the memory of an individual $u$. At moment 0, the distribution of information  pieces could be or the Binomial distribution of parameters $(n, p) = (2^s, 0.5)$, either the  Uniform distribution defined considering a set composed of three pieces of information.  In this work,  $s=5$, establishing the quantity of distinct information on the alphabet $\mathcal{A}$ equal to $2^s = 32$. The Binomial distribution was chosen because it would generate  a relatively large individual empirical entropy, which would facilitate the study of polarization phenomenon. To contrast this situation, an individual can start by choosing between three possible pieces of information uniformly distributed and composed mostly by the symbol 1, to analyze the dynamics when the information start getting polarized. To analyze the dynamics, the results are presented considering both initial distribution.

\subsection{Accepted information}

Since the differences in the number of connections of individuals affects the probability of accepting an information, it is interesting to study, throughout the simulation, how the degree of a vertex can impact on the decision to accept or not information received. For this, it is extracted from the model the total number of accepted information  (i.e., in which the individuals accepted the transmitted information)  for each neighbor nodes of the graph, and then present the values as a function of the degree \(\delta\).

	\begin{figure}[H]
			\centering
			\includegraphics[scale = 0.45]{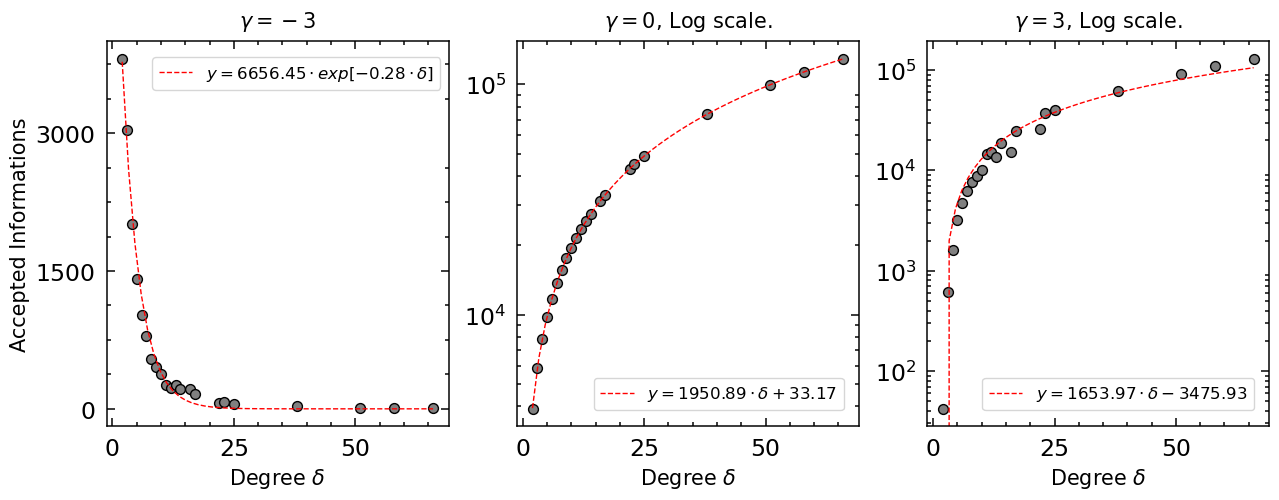}
			\caption{Average number of accepted information and vertex degree.}
			\label{fig:transmissoes_grau1}
		\end{figure}

In the Figure \ref{fig:transmissoes_grau1}, it is observed  the distribution of the number of accepted information. 
For $\gamma=0$ and $\gamma=3$, the relationship between the
number of accepted information and the degree $\delta$ is linear with a smaller line slope for $\gamma =3$ because the most connected individuals are the ones that have the most information accepted. For $\gamma = -3$, the most connected individuals had few or no accepted information, while poorly connected individuals had more accepted information. The relationship between \(y\) and \(\delta\) can also be modeled as a power law \(y = a \cdot e^{b \cdot \delta}\), with \(a > 0\) and \(b < 0\). 

The behavior of the number of accepted information, in relation to the quantities $\gamma$ and $\delta$, is quite similar in simulations with and without polarization.

\subsection{Simulation error}

It will be understood by  \textit{experiment}, a single simulation of a model generated from a defined set of parameters until $T$ iterations starting from a previously generated fixed BA graph. 

Multiple repetitions of an experiment are required to obtain the midpoint of average curves. To ensure robustness, experiments must be repeated a sufficient number of times for smooth midpoint curves of average entropy, average proximity and average polarity when applying in the case.  Arbitrary determination of the number of replications may result in excessive computational costs and insufficient replications could result in poorly smooth curves. The strategy, in this paper, is  to replicate the experiment a sufficient number of times so that the difference between the curves obtained with
a new realization differ little from the average curves computed with the previous simulations. 

Formally, consider a set of parameters defining a model to be simulated over $T$ iterations, and suppose that a number $k$ of replications have been performed, obtaining  a set of curves $\overline{C}^{(1)}, \overline{C}^{(2)}, \ldots, \overline{C}^{(k)}$, where $\overline{C}^{(i)} \in \mathbb{R}^{T}$ for all $i$, $1\leq i\leq k$. With this, it is possible to compute an average curve relative to this set of parameters:
\useshortskip
\begin{equation*}
   \mathcal{C}_k = \dfrac{1}{k} \sum_{i = 1}^{k} \overline{C}^{(i)}.
\end{equation*}

Consider a threshold value $\epsilon >0$. It is performed replications of the experiment until is obtained an $n$ for which the inequality holds:
\useshortskip
\begin{equation}\label{CP}
    \norm{\mathbf{\mathcal{C}}_{n+1} - \mathbf{\mathcal{C}}_{n}}_{2} < \epsilon,
\end{equation}
where \(\norm{\cdot}_2\) is the quadratic norm.

In this work, it is considered $\epsilon = 10^{-8}$ and $\overline{C}=\overline{H}$, $\overline{C}=\overline{S}$ or $\overline{C}=\overline{\pi}$ when is simulated the average entropy,  average proximity or the average polarity respectively.

\subsection{Simulation scenarios} 

The dynamics of the model will be analyzed based on the evolution curves of the quantities: average entropy $\overline{H}$, average proximity  $\overline{\mathcal{S}}$, average polarity $\overline{\pi}$ (in cases where there is polarization) and the probability distribution of the binary codes. The behavior of these quantities will be observed considering the variation of the parameters $\kappa$, $\gamma$, $\alpha$, and $\omega$.

There are four scenarios to be analyzed: (i) the population is not polarized ($\alpha=\omega=0$); (ii) the population is divided into a neutral and a positive or negative polarized 
group ($\alpha=0$ and $\omega>0$ or $\alpha>0$ and $\omega=0$); (iii) the population is composed by a neutral group and two polarized groups ($0<\alpha + \omega<1$); and (iv) the population is divided into two polarized groups ($\alpha + \omega=1$). 

For the purpose of defining transition probabilities, consider the following notation
\begin{equation}\label{eq:transition_prob}
    \begin{array}{ll}   
  P(y = 0 | x = 0) = P_{00}, & P(y = 1 | x = 0) = P_{01}, \\
  P(y = 0 | x = 1) = P_{10},  & P(y = 1 | x = 1) = P_{11}.
\end{array}
\end{equation}

\subsubsection{The population is not polarized ($\alpha=\omega=0$)}

In this case, all individuals are in $\overline{V}$. Using the notation in \eqref{eq:transition_prob}, the transition probabilities are defined as
\begin{equation}\label{eq:not_polarizated}
    \begin{array}{ll}   
P_{00} = 1 - \delta_u, &  P_{01}=\delta_u, \\
P_{10} =  \delta_u,  &  P_{11}  = 1 - \delta_u,
\end{array}
\end{equation}
where $\delta_u$ defined by \eqref{delta}.

\begin{figure}[H]
			\centering
			\includegraphics[scale = 0.47]{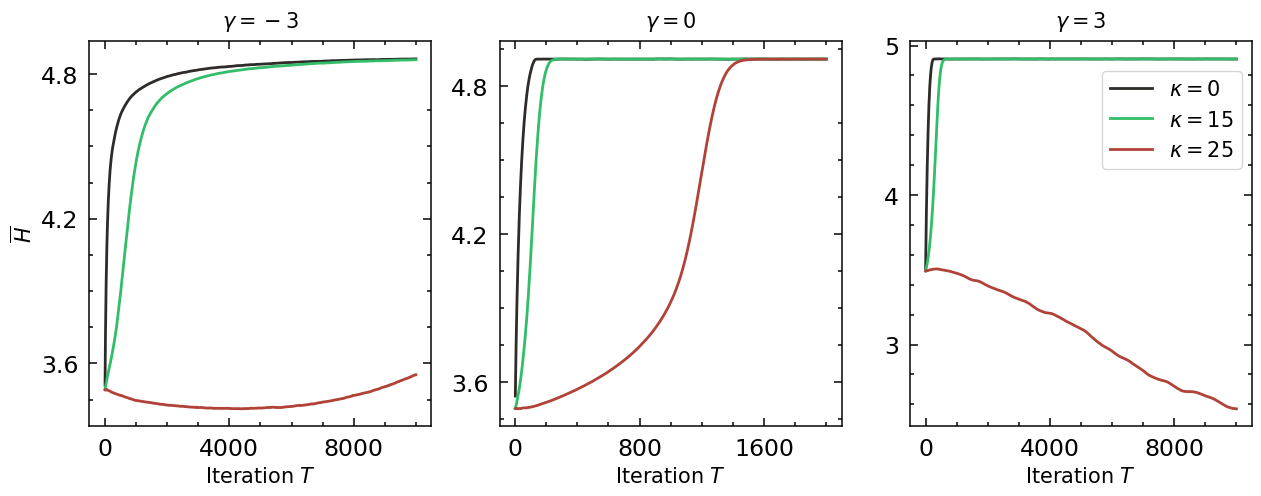}
			\caption{Temporal evolution of average entropy as a function of the parameters \(\kappa\) e \(\gamma\). }\label{fig:curvas_evolucao_entropia_barabasi}
		\end{figure}

In the Figure \ref{fig:curvas_evolucao_entropia_barabasi} is shown the evolution curves of average entropy for several values of $\kappa$ and $\gamma$.
When $\gamma=0$, regardless of the value of $\kappa$, entropy converges to its maximum value. This means that distortion occurs, and all information are available on the memories. 
When $\gamma=\pm 3$, entropy converges to its maximum value whenever $\kappa=0$ and $15$. The difference between these cases is the speed of this convergence, now reduced due to social preferences. When $\kappa=25$, the population has the tendency to not disturb information and therefore, the information varies little and different information takes a while to appear and be disseminated.

\begin{figure}[H]
			\centering
			\includegraphics[scale = 0.47]{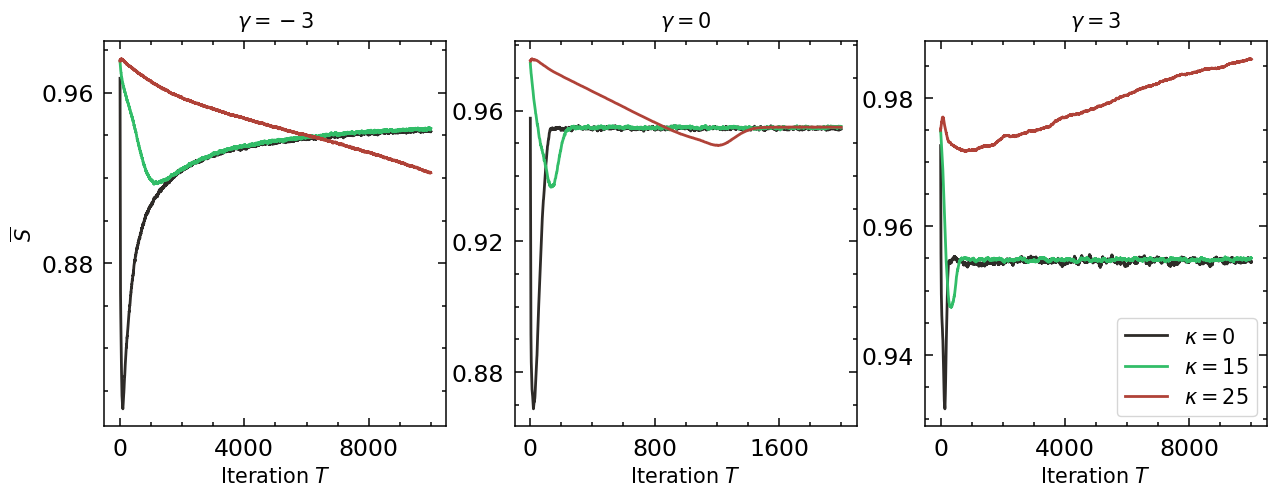}
		\caption{Time evolution of the average proximity as a function of the parameters \(\kappa\) and $\gamma$.}
			\label{fig:curvas_evolucao_proximidade_barabasi}
		\end{figure}

In the Figure \ref{fig:curvas_evolucao_proximidade_barabasi} it is shown the average proximity for several values of $\kappa$ and $\gamma$.

Except for the cases $\kappa= 25$ with $\gamma=\pm 3$,  the curves start with a fall, as pronounced as the value of $\kappa$ is smaller, followed by an increase in proximity until it shows a tendency to stabilize. The initial fall happens due to rapidly increasing entropy (see Figure \ref{fig:curvas_evolucao_entropia_barabasi}) where new information appears in the population, resulting in temporary differences among neighbors. This decline occurs to the point where all information circulates among the population and there is no way for individuals to diverge ideologically. From here, the tendency is to homogenize the distribution of information. At this stage, the proximity curve begins to grow again, showing a tendency toward stabilization. 

For the case  $\kappa= 25$ and $\gamma=-3$, it is a natural presumption that after a longer time than simulated, all possible information will be available in the memory of individuals which will cause the curve for average entropy be ascending (as it begins to appear in the Figure \ref{fig:curvas_evolucao_entropia_barabasi}) and both, this curve and the average proximity curve, converge to the same previous values as $\kappa=0$ and $15$ (see also Figure \ref{fig:curvas_evolucao_proximidade_barabasi}). The slow evolution of both curves can be explained by the fact that most interactions happen among unpopular individuals. Thus, new information takes a long time to be disseminated into the population. 

For $\kappa= 25$ and $\gamma=3$, however, the proximity does not show the same behavior as the previous cases, increasing instead. The parameter $\gamma = 3$ causes a generalized preference toward a well-connected individuals. Thus, new information presented by these individuals might prevail in the population. In addition, the acceptance probability for popular indivi\-duals converges toward $1$ as the proximity increases, while less connected individuals remain penali\-zed by the relative popularity factor. Therefore, the preference toward the highly connected individuals is reinforced as the proximity increases. In this scenario, it can be assumed that
the global proximity will increase towards 1, while the entropy will stabilize at a non-zero value, corresponding to the final, homogeneous state of the population.

The previous conclusions were obtained starting with the information in the memories modeled by a Binomial distribution, but what if the dynamics of rumor dissemination begins with three equally probable pieces of information?

\begin{figure}[H]
			\centering
			\includegraphics[scale = 0.50]{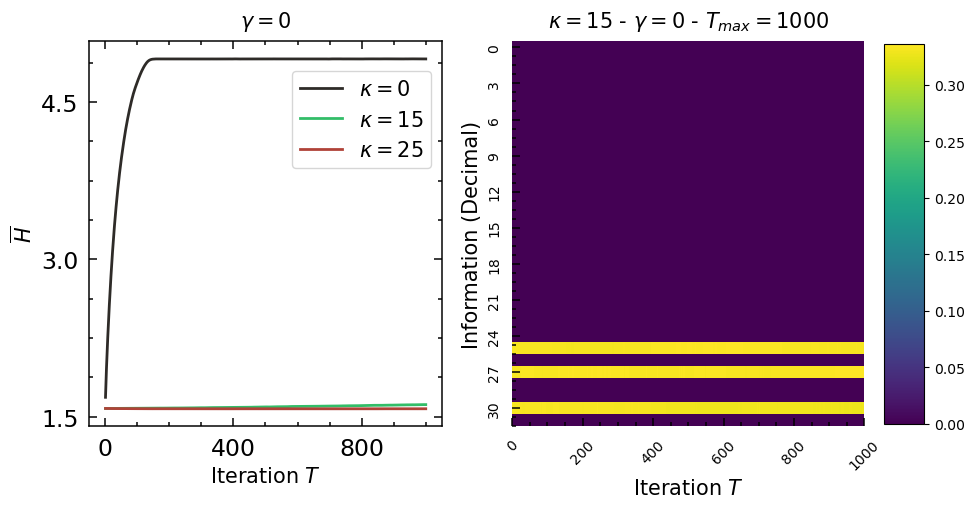}
			\caption{Temporal evolution of the average entropy with memories initialized from a Uniform distribution over a set of three pieces of information.} 
                \label{fig:no-polarization-unifom}
		\end{figure}

In fact, in Figure \ref{fig:no-polarization-unifom} it is present the average entropy obtained from an initial distribution containing only three pieces of information composed, for the most part, by 1 (positive polarized information). It is noted that the average entropy converges to the maximum value when $\kappa=0$. This occurs because, even starting with three equally probable pieces of information, eventually, all the information is circulating in the individual's memories, something that does not happen with $\kappa=15$ or $\kappa=25$. In those cases, the initial average entropy remains low because, for increasing it is necessary a greater volume of accumulated distortions and, therefore, new information circulating into the population, but with those $\kappa$ values, the probability of that happening is strongly reduced.

\subsubsection{The population is divided into a neutral and a positive or negative polarized  group ($\alpha=0$ and $\omega>0$ or $\alpha>0$ and $\omega=0$)}
It will be analyzed the behavior of neutral and positive polarized group (similarly for negative polarized group).  To study under which conditions positive polarized individuals can bias the information over the neutral population, it is necessary to define the following transition probabilities based on the notation \eqref{eq:transition_prob}.
If an individual $u\in \overline{V}$, the transition probabilities were described in \eqref{eq:not_polarizated}. If  an individual $u\in V_{\uparrow}$ the probabilities are given by
\begin{equation}\label{eq:positive_polarizated}
    \begin{array}{ll}   
P_{00} = 1 - \delta_u - \xi_u, &  P_{01}=\delta_u + \xi_u, \\
P_{10} =  \delta_u,  &  P_{11}  = 1 - \delta_u,
\end{array}
\end{equation}
where $\delta_u$ was given by \eqref{delta} and $ \xi_u$ by \eqref{xi}. Note that these probabilities are biased to change 0 by 1, i.e. distortion of the form \(0 \to 1\).

It is highlighted that the transition probabilities defined by \eqref{eq:positive_polarizated} are all valid since $\delta_u + \xi_u \leq 0.5$ achieving equality when   \(\lambda =\kappa = 0\). 
    \begin{itemize} 
    \item When $\kappa=0$: If $\lambda=0$, the transition probabilities for $u\in \overline{V}$ are 
	\begin{equation*}
				P_{01} = P_{00} = P_{11} = P_{10} = 1/2,
			\end{equation*}
and for an individual $u\in V_{\uparrow}$ the probabilities are 
 \begin{equation*}
    \begin{array}{ll} 
P_{00} = 0, &  P_{01}=1, \\
P_{10} =  \frac{1}{2},  &  P_{11}  =\frac{1}{2}.
\end{array}
\end{equation*}

Polarized individuals will change 0 to 1 with probability 1,  inducing the accumulation of positive polarized information, that is, information that contains more 1 than 0. However, since both groups, neutral and polarized, have a probability of changing 1 to 0 in the information to be transmitted, the population will have a great variety of information available, which influences the behavior of the average entropy which does not converge to its maximum.  Keeping \(\kappa = 0\) and choosing \(\lambda > 0\), the transition probabilities for the polarized group will now be given by
\begin{center}
     $P_{01} = 1/2 + \xi_u$, $P_{00} = 1/2 - \xi_u$, $P_{11} = P_{10} = 1/2$.
\end{center}

In this case, as it can be seen in the Figure \ref{fig:polarized_1} with $\lambda=1$, the behavior of the average entropy is similar to the previous case when $\lambda=0$. It is possible to show that, when \(\lambda\) increases,  the probability \(\xi_{u}\) decreases  and individuals will be less prone to remove bits \(0\) from the information. Regardless of the value of \(\alpha\), the polarized group loses its influence over the information distribution and the average entropy behavior becomes similar. 
			\begin{figure}[H]
				\centering
				\includegraphics[scale = 0.5]{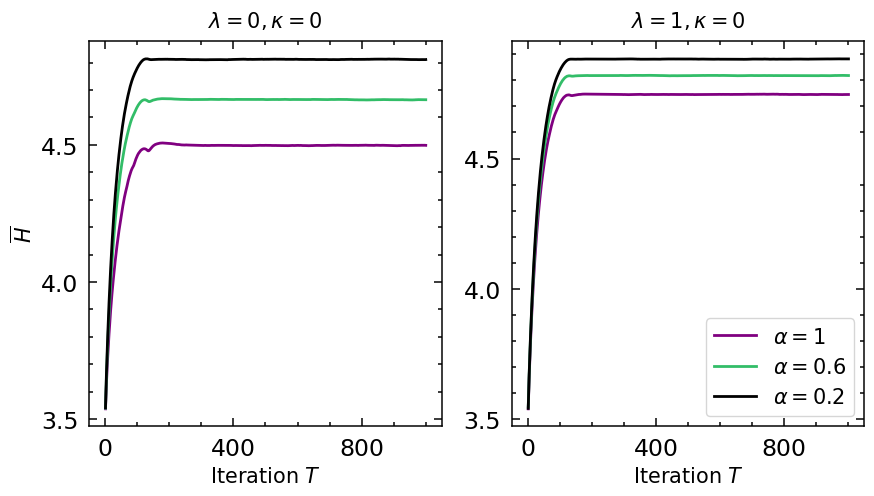}
				\caption{Temporal evolution of average entropy as a function of the parameters \(\lambda\) and \(\kappa=0\).}
				\label{fig:polarized_1}
			\end{figure}
		
\item When $\kappa>0$: If \(\lambda = 0\), the transition probabilities for the polarized group will be given by
\begin{center}
    \(P_{01} = \delta_u + 1/2\), \(P_{00} = 1/2 - \delta_u\), \(P_{11} = 1 - \delta_u\), \(P_{10} = \delta_u\).
\end{center}

In this case, the polarized group still has a tendency of removing \(0\) bits from the information before transmitted, and a higher chance of preserving bits \(1\). As \(\kappa\) increases, \(\delta_{u} \to 0\), and the polarized group will preserve polarized information ($P_{11} \to 1$, $P_{10} \to 0$). In this context, the neutral group is less prone to information distortion, transmitting more faithfully the information received in the past. As a result, the preservation effort of the polarized group is enough to totally bias the population (Figure \ref{fig:polarized_2}). 
			\begin{figure}[H]
				\centering
				\includegraphics[scale = 0.5]{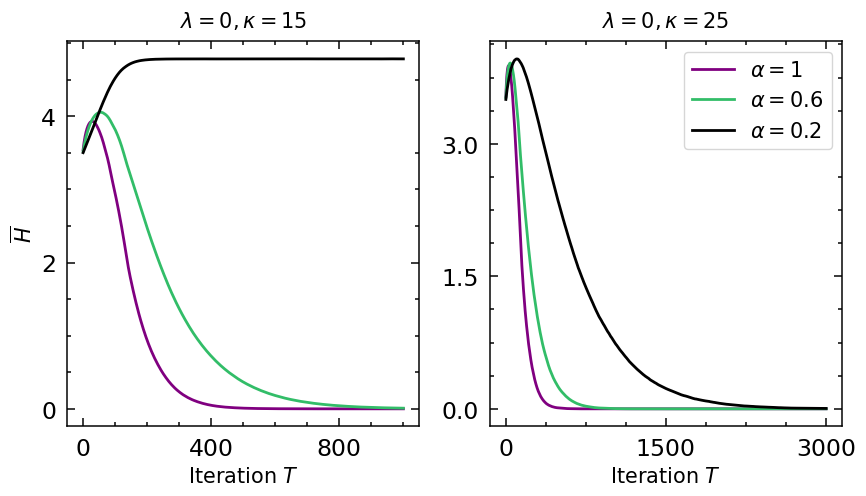}
				\caption{Temporal evolution of average entropy as a function of the parameters \(\lambda=0\) e \(\kappa>0\).}
				\label{fig:polarized_2}
			\end{figure}
		
 With $\lambda>0$, the transition probabilities for a positive polarized individual will be those described in the equation \eqref{eq:positive_polarizated}. Specifically, if $\lambda=1$ the probability of positive distortion is lower and different from the previous case. With $\kappa=15$ and $\alpha=0.2$ or $\alpha=0.6$, the neutral group manages to transmit information with more zeros, which causes the entropy to increase. The same difference with the same reason occurs in the case $\kappa=25$ and $\alpha=0.2$. For case $\kappa=15$ and $\kappa=25$ with $\alpha=1$, the Figure \ref{fig:polarized_2_lambda_1} shows that, although the dynamic started with the initial Binomial distribution, the information that circulates was polarized, which causes the average entropy to decrease.
\begin{figure}[H]
				\centering
				\includegraphics[scale = 0.5]{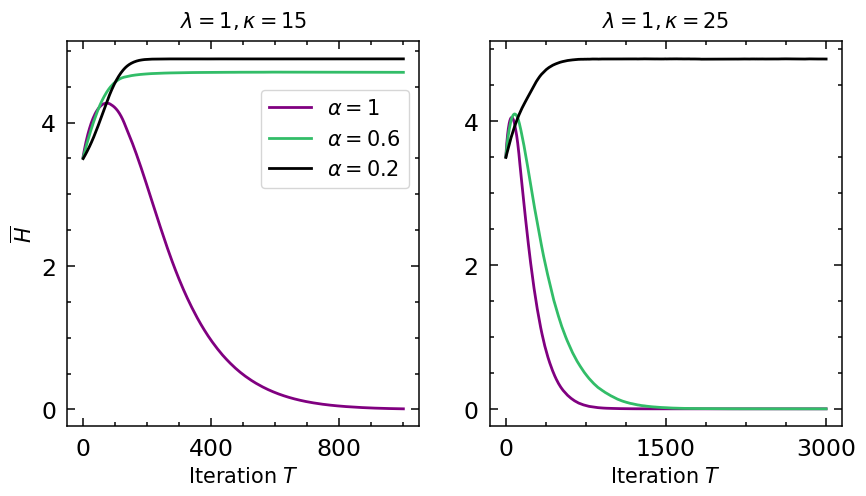}
				\caption{Temporal evolution of average entropy as a function of the parameters \(\lambda=0\) e \(\kappa>0\).}
				\label{fig:polarized_2_lambda_1}
			\end{figure}
		
What would happen if the dynamics begins with an already polarized distribution? Suppose the dynamic begins with three pieces of information, uniformly distributed, where most of the elements are 1 (positive polarized information). That implies low initial entropy.  In the Figure \ref{fig:uniform_polarized}, it is possible to observe that for  \(\kappa = 15\), even though $\alpha=0.2$, the average entropy decrease until its minimum value. When $\alpha< 1 $, neutral individuals become an echo chamber just passing on the biased information produced by the polarized group. At the end of the dynamic, there is only one piece of information that circulates: 11111. 
\begin{figure}[H]
    \centering
    \includegraphics[scale = 0.45]{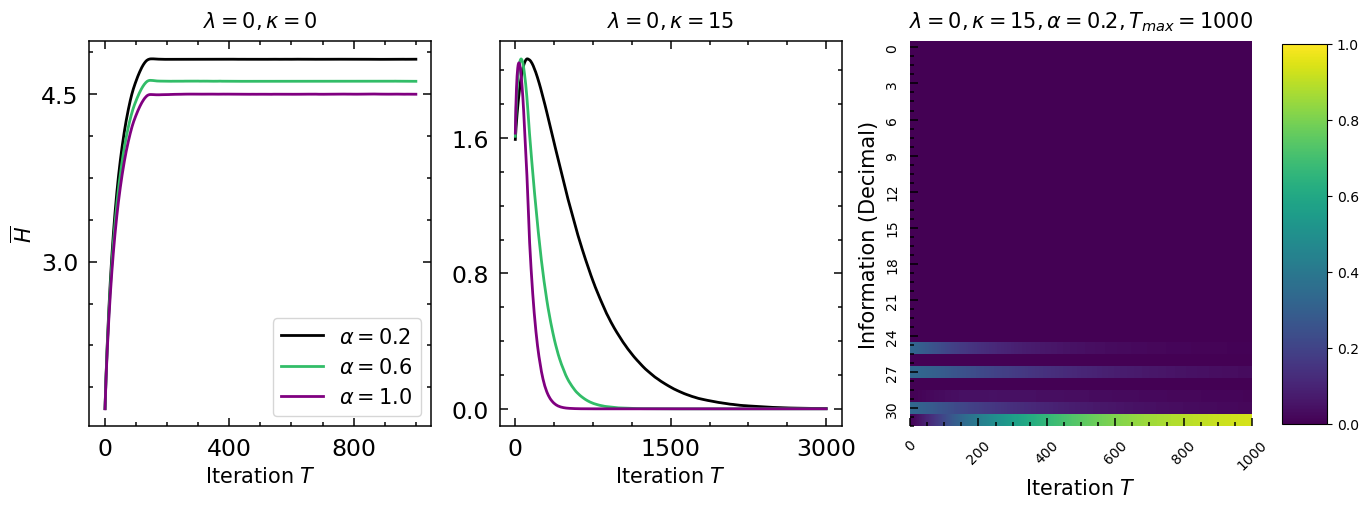}
    \caption{Average entropy with initial Uniform distribution using $\lambda=0$.}
    \label{fig:uniform_polarized}
\end{figure}

When $\kappa=25$ the behavior of curve is the same as with $\kappa=15$ (Figure \ref{fig:uniform_polarized}) but the speed  of convergence is lower. Due to the similarity of the cases, the graphs were omitted.
 
\end{itemize}

\subsubsection{The population is composed by a neutral group and two polarized groups ($0<\alpha + \omega<1$)}
\begin{figure}[H]
				\centering
				\includegraphics[scale = 0.50]{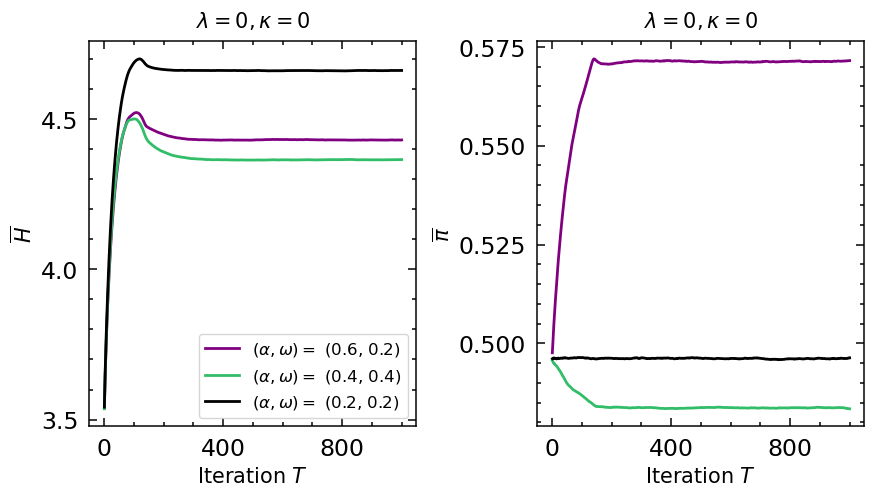}
				\caption{Average entropy and polarity for positive and negative polarized individuals with \(\kappa = \lambda = 0\).}
				\label{fig:polarized_3}
	\end{figure}

Consider \(\kappa = \lambda = 0\). When the number of polarized individuals is equal in each group, entropy increases as the size of the neutral group increases. When a polarized group has more individuals than the other, the entropy increases compared to the balanced polarized groups, although the percentage of neutral individuals be the same (for instance, in the Figure \ref{fig:polarized_3} it is possible to compare the curves with $(0.6, 0.2)$ and $(0.4, 0.4)$). This means that, although the positive polarized group is the majority and therefore there is a higher amount of positive polarized information, that information is subject to distortions caused by the negative polarized group even though this is a group with fewer individuals.  
By analyzing the average polarity curve, it is possible to see that for unbalanced  polarized groups, it is greater than for balanced polarized groups. For instance, the average polarity obtained with $(\alpha, \omega) = (0.6, 0.2)$ is higher than the one found for $(\alpha, \omega) = (0.4, 0.4)$,  showing the prevalence of positive polarized information within the population. Although the Figure \ref{fig:polarized_3} shows the average polarity lower for $(\alpha, \omega) = (0.4, 0.4)$ than for $(\alpha, \omega) = (0.2, 0.2)$, both are close to 0.5, which is expected for balanced groups. 

\begin{figure}[H]
				\centering
				\includegraphics[scale = 0.50]{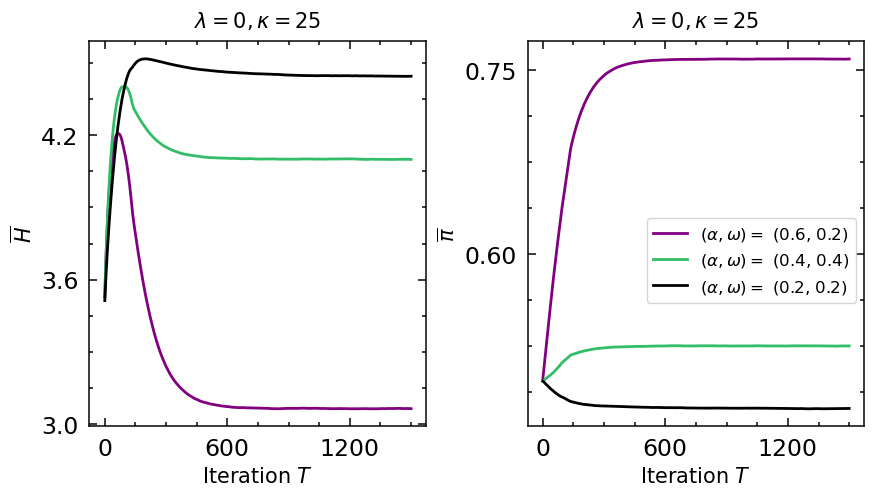}
				\caption{Average entropy and polarity for positive and negative polarized individuals with \(\lambda = 0, \kappa = 25\).}
				\label{fig:polarized_4}
			\end{figure}

As \(\kappa\) increases, the positive polarized (similarly for negative polarized) individuals act less in the removal of \(0\) bits (\(1\) bits, respectively) and more in the preservation of \(1\) bits. (\(0\), respectively). The polarized groups act accumulating polarized information relative to its bias, and the general tendency for the population will be determined by the majority polarized group. In this sense, the average entropy decreases inasmuch the proportion of polarized individuals increases and also as the proportion of one of the groups overcome the others (Figure \ref{fig:polarized_4} with $\kappa=25$). Therefore, if the majority group is the positive polarized one, then the lower average entropy corresponds to a higher polarity. Nonetheless, it is not possible that the entire population will be biased towards one direction, given that the polarized group of opposite polarity preserves information from its polarity, and modifies incoming information from the other polarity.
		
\begin{figure}[H]
				\centering
				\includegraphics[scale = 0.50]{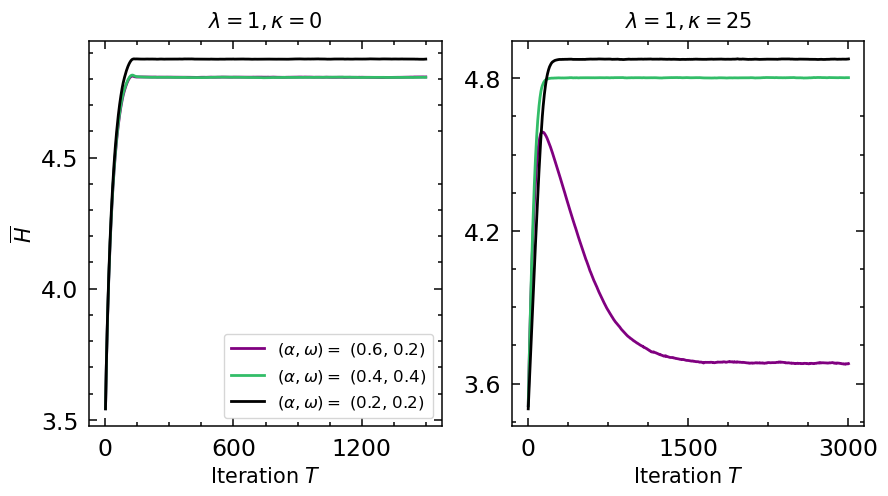}
				\caption{Average entropy and polarity for positive and negative polarized individuals with \(\lambda = 1, \kappa = 0\) and \(\kappa = 25\)}
				\label{fig:polarized_5}
			\end{figure}

The increase in the parameter \(\lambda\) reduces the distortion bias from polarized individuals, making it more difficult for the majority polarized group to bias the population (Figure \ref{fig:polarized_5}). In this scenario, the average entropy for the case $\lambda=0, \kappa=25$ is lower than the case $\lambda=1, \kappa=25$, because there  is a low concentration of polarized information.

The parameter \(\kappa\) is a determinant factor to enable polarization effects. This is because a greater information variety is a hindrance to the influence of polarized groups over the population, and for higher values of \(\kappa\), the neutral group simply reproduces the information it receives, that is, essentially reproducing the information circulation pattern in the population. As it was shown previously, this behavior can be observed not only by increasing \(\kappa\), but also by initializing the population with an information distribution already less diversified. It was observed in the previous section that this makes the population more susceptible to being biased by the polarized group (see Figure \ref{fig:polarized_5}).

\begin{figure}[H]
				\centering
				\includegraphics[scale = 0.50]{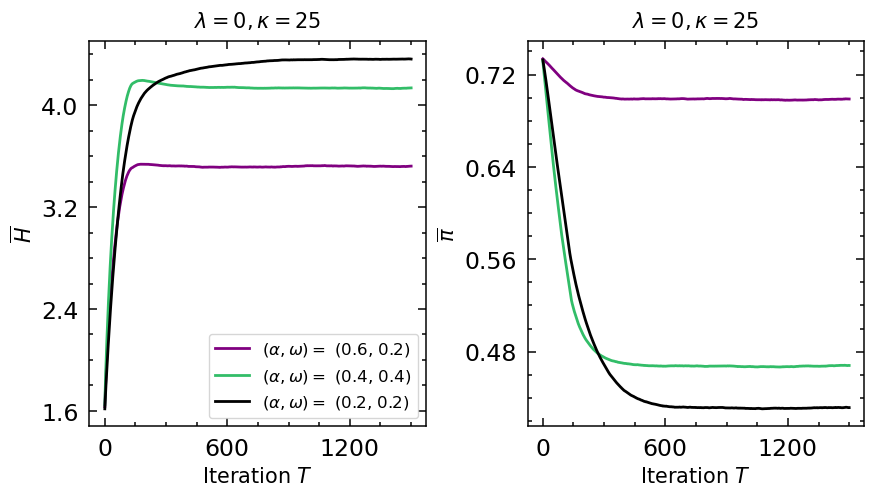}
				\caption{Average entropy and polarity for positive and negative polarized for initial Uniform distribution with \(\lambda = 0, \kappa = 25\).}
				\label{fig:polarized_7}
			\end{figure}

Now consider that the dynamics of information transmission begins with a Uniform distribution. In Figure \ref{fig:polarized_7}, it is possible to observe the evolution curves for the average entropy and average polarity obtained starting the transmission with only three pieces of positive biased information. It can be noticed in the figure that the convergence values for the average entropy and average polarity are close to the ones observed in Figure \ref{fig:polarized_4}. Being more precise, for \((\alpha, \omega) = (0.6, 0.2)\), the average entropy's convergence value is \(3.22419 \pm 0.00030\) and the average polarity's convergence value is \(0.73852 \pm 0.00005\). 
In the present case, those numbers are \(3.46689 \pm 0.00101\) and \(0.70150 \pm 0.00015\), respectively. Therefore, the final state of the information transmission  appears to be independent of the initial information distribution.

As a conclusion,  the information transmission depends entirely on the behavior of the polarized groups, and its proportions. The difference between this scenario and the previous one is that previously there was a single polarized group, therefore the neutral group, by reproducing the general dynamics, amplified the effect of the polarized group. However, in the present case, what impedes one group from dominating the population is the presence of the other polarized group.

\subsubsection{The population is composed by two polarized groups ($\alpha + \omega=1$)}
As presented in the previous section, if two groups with opposite polarities are present in the population, the information transmission dynamic is determined by the interaction between these two groups. 

\begin{figure}[H]
				\centering
				\includegraphics[scale = 0.50]{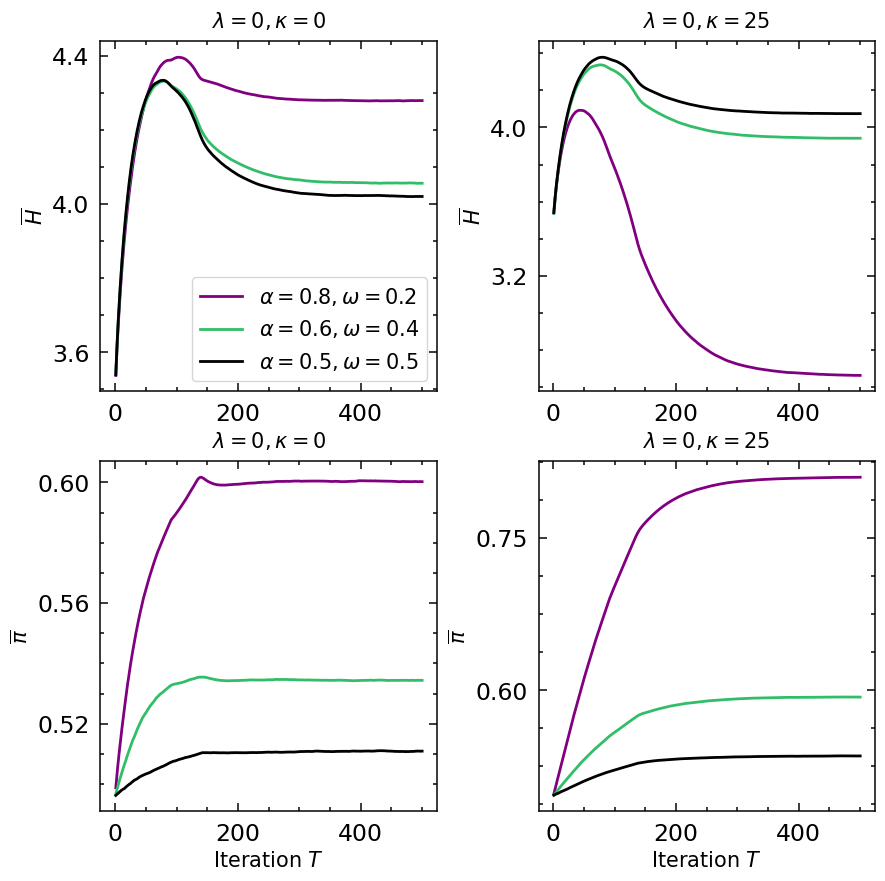}
				\caption{Average entropy and polarity for $\lambda=0$.}
				\label{fig:polarized_6}
			\end{figure}

It is noted that there is an association between low average entropy and a higher average polarity due to the majority group's bias. It is observed, however, that this does not happen when \(\kappa = 0\) (Figure \ref{fig:polarized_6}).  If \(\kappa = 0\), with probability 1, the positive polarized group introduces bits 1 in the information before being transmitted and the negative group removes it, therefore if the number of individuals in each group is similar, information will accumulate at the poles and in this manner, the decrease in average entropy is accompanied by an average polarity closer to 0.5.

\begin{figure}[H]
				\centering
				\includegraphics[scale = 0.50]{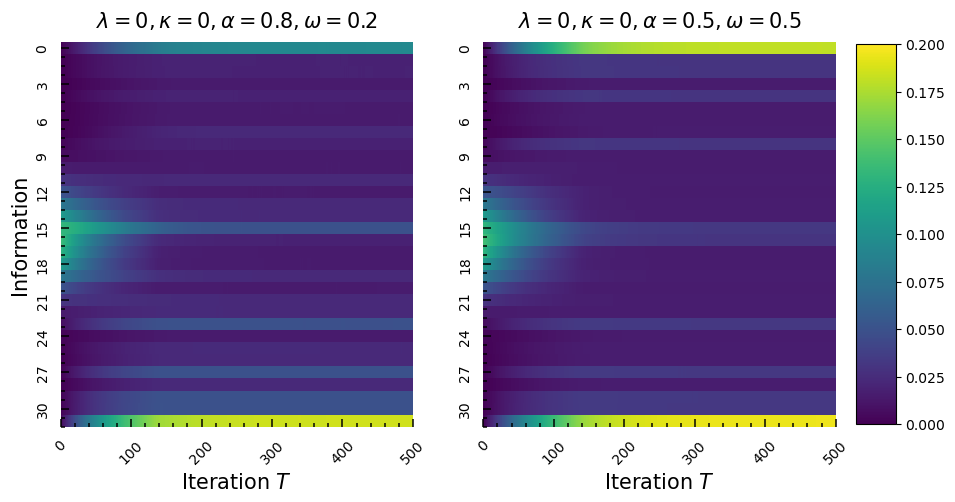}
				\caption{Information probability distribution.}
				\label{fig:polarized_distribution_1}
			\end{figure}

 Making \(\kappa > 0\) such that \(\delta_{u} \to 0\), the positive polarized group acts instead to preserve the bits 1 (respectively, bits 0 for the negative polarized group); therefore the majority group biases the population toward the accumulation of information at its corresponding pole. For $\kappa=25$, the Figure \ref{fig:polarized_6} shows the average entropy decaying, however the population is not entirely biased due to the presence of the group with opposite polarity, which with probability \(1 / 2\) introduces biased information in its favor, and with probability \(1\) preserves information from its polarity.
Figure \ref{fig:polarized_distribution_1} shows that, at the end of the simulation, when a polarized group is in majority (specifically \((\alpha, \omega) = (0.8, 0.2)\)) the information that prevails is that of this group (in this case, the most likely piece information is 11111). When the polarized groups are balanced, the most likely pieces of information are 00000 and 11111.

\begin{figure}[H]
				\centering
				\includegraphics[scale = 0.50]{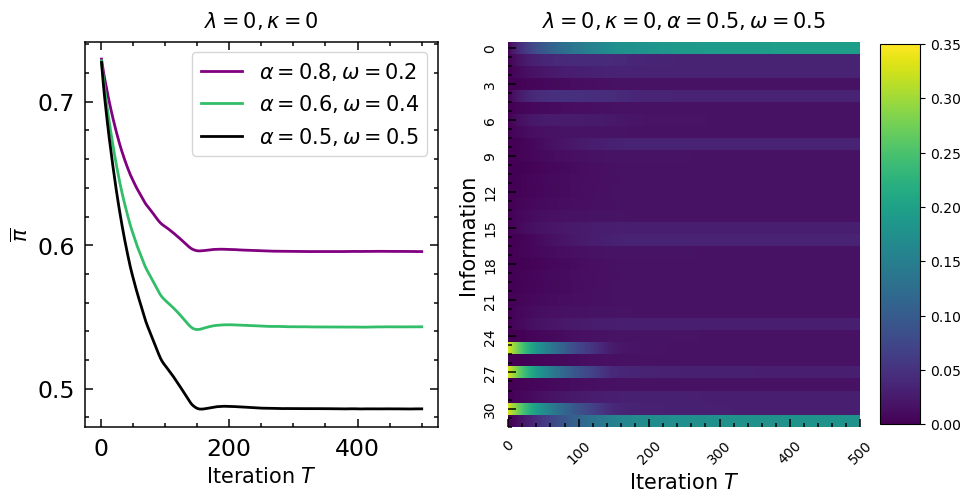}
				\caption{Average polarity and information for initial Uniform distribution}
				\label{fig:uniform_two_polarized_groups}
			\end{figure}

The fact that the transition probability for polarized individuals does not depend solely on \(\delta_{u}\), but also on \(\xi_{u}\), starting information transmission using a Uniform distribution with only three pieces of information (in this case positive polarized) in the individual's memory will not change the overall dynamics of the transmission, as it was described until now. If, as mentioned, it is started with three pieces of information composed mostly of 1 (positive polarized information), this will not imply an advantage for the positive group. In fact, for $\lambda= \kappa=0$, it is observed a decrease in the average polarity toward a value close to the one observed in Figure \ref{fig:polarized_6}  as the negative polarized individuals act producing or preserving information with lower polarity (Figure \ref{fig:uniform_two_polarized_groups}). Similarly, starting the population with a negative polarized distribution will not give an advantage to the negative group, in counterpoint to its disadvantage regarding its number of individuals.
			
\section{Discussion}\label{S:conclusion}

In this study, it is analyzed the dynamics of information exchange in a social network modeled by a BA graph. The proposed model considered individual memory, ideological proximity between individuals and distortion influenced by  polarization.  To achieve this, it is modeled the information transmitted  as binary codes in such a way that, by including an individual memory of received information, it was possible to define a probability distribution of information for each individual. This allowed to apply metrics based on entropy to analyze the evolution of the information distribution contained into a population's memories, and measure the similarity of information between individuals using the Jensen-Shannon divergence measure. Several scenarios to transmit the information were examined, including one without polarization bias, as well as considering polarization bias with different proportions of polarized individuals in the population.

When the transmission of information is studied without polarized individuals and starting with the Binomial distribution in their memories, it is observed that, for $\gamma=0$ the average entropy tends to reach its maximum value because the initial distribution converges to the Uniform distribution. This indicates that everyone's memory contains similar information and in similar proportions. For $\gamma\neq 0$ and \(\kappa \geq 15\) happens as in the previous case only the speed of convergence is lower. 
When $\kappa=25$ and $\gamma = -3$, it is presumed that  all available information will be in the memories at some point and the initial Binomial distribution will converge to the Uniform distribution, i.e., the average entropy will converge, eventually, to its maximum value, taking a longer time by the fact that information is disseminated primarily by unpopular individuals. For $\kappa = 25$ and $\gamma = 3$, very well-connected individuals, now preferred by the population, easily disseminate information they produce and which then prevail. In this scenario, it can be presumed instead that the information distribution will stabilize in a non-Uniform distribution, determined by the well-connected individuals, as the average proximity increases toward $1$.

When the initial Binomial distribution is changed to the Uniform distribution between three pieces of positive polarized information, it is noted that the average entropy converges to the maximum value when $\kappa=0$, but remains practically unchanged, until the end of the simulation, for $\kappa = 15$ and $25$. This occurs because, even starting with three equally probable pieces of information, eventually, for $\kappa=0$, all the information is circulating in the individual’s memories since there is no distortion control and it is inevitable that new information emerge. This  not happen with $\kappa \geq 15$, contrarily to what was observed when starting with the Binomial distribution. Therefore, a population with little diversity of opinions will naturally tend to keep them in memory and with few variations for a longer period of time.

In scenarios where the population includes a polarized group, it is observed that it is less susceptible to being biased by the polarized group if there is no strong control of distortions ($\kappa=0$). However, for a higher value of \(\kappa\), e.g. \(\kappa = 25\), the population is biased by the polarized group, even if it is a minority. An initial distribution of information with low entropy, (Uniform distribution), also makes the population susceptible to bias as the low variety of initial information also hinders the emergence of new information in the population. As a result, the neutral population acts as an echo chamber for the polarized group, which introduces new information in favor of its bias. This phenomenon is not observed when there are two groups with opposite polarities in the population. The neutral group reproduces the information that they listen, but without a preference for either pole. The dynamic of information transmission  is mainly determined by the proportion of each polarized group. Such a group tends to produce or preserve information from its pole, depending on the value of \(\kappa\), which prevents the population from being completely biased towards one information pole.

In summary, the model shows that a population is more susceptible to the influence of a biased group, if the it does not have to compete with another group of opposite polarity and if the population has a low production of different information. This low production could occur either due to an initially homogeneous distribution of information or to the low production of new information by high control distortion ($\kappa\geq 15$). The preference for information from more connected individuals ($\gamma>0$) lead the population to accept it readily resulting in a low average entropy.

In all these scenarios, it is possible to interpret that the majority of the neutral population acts passively, receiving information from a specific group and then, passing it on to their neighbors. As seen in the case where the it is subject to the influence of a single polarized group, such passive behavior transforms it into an echo chamber, and the polarized group can easily bias the entire population.



\end{document}